# Barrier height prediction by machine learning correction of semiempirical calculations

Xabier García-Andrade,[a,b] Pablo García Tahoces[b], Jesús Pérez-Ríos[c,d] and Emilio Martínez Núñez*[a]

[a]Department of Physical Chemistry, University of Santiago de Compostela 15782, Spain

[b]Department of Electronics and Computer Science, University of Santiago de Compostela 15782, Spain

[c]Department of Physics, Stony Brook University, Stony Brook, New York 11794, USA

[d]Institute for Advanced Computational Science, Stony Brook University, Stony Brook, NY 11794-3800, USA

**ABSTRACT**

Different machine learning (ML) models are proposed in the present work to predict DFT-quality barrier heights (BHs) from semiempirical quantum-mechanical (SQM) calculations. The ML models include multi-task deep neural network, gradient boosted trees by means of the XGBoost interface, and Gaussian process regression. The obtained mean absolute errors (MAEs) are similar or slightly better than previous models considering the same number of data points. Unlike other ML models employed to predict BHs, entropic effects are included, which enables the prediction of rate constants at different temperatures. The ML corrections proposed in this paper could be useful for rapid screening of the large reaction networks that appear in Combustion Chemistry or in Astrochemistry. Finally, our results show that 70% of the bespoke predictors are amongst the features with the highest impact on model output. This custom-made set of predictors could be employed by future Δ-ML models to improve the quantitative prediction of other reaction properties.

## 1. Introduction

Transition state theory (TST) provides an useful means to study the kinetics of elementary chemical reactions.[1] Depending on the specific version, TST requires a more or less exhaustive knowledge of the potential energy surface of the system.[2] In the absence of strong tunnelling effects, the value of the Gibbs energy of activation $\Delta G^{\ddagger}$ (Gibbs energy difference between the transition state (TS) and the reactant) is sufficient to predict the rate of reaction. At 0 K, $\Delta G^{\ddagger}$ is just the electronic energy difference between the TS and reactant including their zero-point vibrational energies (ZPEs), called the barrier height (BH). Although the BH does not include the thermal correction to enthalpy and the entropic contribution, sometimes it is employed as a proxy for the true Gibbs energy of activation. Nevertheless, predicting highly accurate BHs (of sub kcal/mol accuracy) requires the use of expensive *ab initio* methods, such as the gold standard coupled cluster including single and double excitations with perturbative triple excitations [CCSD(T)].[3] Fortunately, today's state-of-the-art density functionals predict BHs that are rather close to the accurate CCSD(T),[4] thus being the method of choice for modelling large systems. However, even DFT becomes prohibitive for biochemical systems or for complex reaction networks of medium-size systems.

With the surge of large computational and experimental datasets, machine learning (ML) is shifting the paradigm to data-driven predictive modelling. This approach has been pursued to predict activation energies and BHs in previous works.[5-15] By way of example, Choi et al. developed different ML models to predict activation energies of gas-phase reactions, with the tree boosting method showing the best performance.[5] More recently, Green and co-workers have demonstrated that it is possible to predict accurate BHs using a deep learning (DL) model given only reactant and product graphs.[6, 8] Green's DL model was trained on a gas-phase organic chemistry (GPOC) data set of 12,000 chemical reactions involving carbon, hydrogen, nitrogen and oxygen. The calculations were carried out at the DFT ωB97X-D3/def2-TZVP quantum chemistry level, which has shown to predict BHs with errors of only 2.28 kcal/mol.[4] They recently improved the model using fewer parameters and proper data splits to estimate performance on unseen reactions.[8] Also, Habershon and co-workers employed this basis set to predict rates of chemical reactions.[16] Alexandrova and co-workers have also shown that topological descriptors of the quantum mechanical charge density in the reactant state can be used to predict BHs for Diels-Alder reactions.[9] Hybrid models combining traditional TS modelling and ML are also employed to predict BHs for nucleophilic aromatic substitution reactions in solution.[10]

Semi-empirical quantum mechanical (SQM) methods are significantly faster than DFT and provide results with sufficient accuracy when applied to molecules of the same type as those of the training set.[17] However, except when the interest is in a specific reaction,[18-22] training sets do not usually include data of TSs, which results in inaccurate BH predictions. In an attempt to model the reactivity of organic reactions with useful accuracy, Stewart developed the SQM method called PM7-TS.[17] Using a training set of 97 BHs obtained from collections of high-level calculations, the mean absolute error (MAE) using PM7-TS was 3.8 kcal/mol, as compared with the MAEs for PM7 of 11.0 kcal/mol and for PM6 of 12.2 kcal/mol.[17] However, Jensen and co-workers benchmarked PM7-TS using barrier heights for five model enzymes and found a MAE of 19 kcal/mol, while the MAEs for PM6 and PM7 were around 12-15 kcal/mol.[23] Iron and Janes[24] have also shown that SQM methods perform very poorly predicting transition metal barrier heights: using a new dataset with high-accurate energies, the MAEs of PM6, PM7 and PM7-TS are 21.6, 106.4 and 68.2 kcal/mol. In his PM7 paper, Stewart already acknowledged that the predictive power of PM7-TS was unknown at the time, and suggested parameter re-optimization as more BHs became available.[17]

An alternative to parameter optimization is to develop analytical[25] or ML corrections of the SQM calculations. The latter are usually termed Δ-machine learning because the model predicts the difference between the benchmark and the approximate baseline calculation (SQM in this case).[26] There are some examples in the literature of the successful use of machine learning to improve the accuracy of both DFT[27-29] and SQM calculations.[30, 31]

In this work we leverage machine learning to predict barrier heights with DFT accuracy at the cost of SQM calculations. The model employs multi-task deep neural network (DNN), gradient boosted trees by means of the XGBoost interface,[32] and Gaussian process regression (GPR) trained on a curated version of the GPOC data set.[7] Gradient boosting regression has been successfully applied to predict BHs in Diels-Alder reactions,[9] and the reactivity of transition metal complexes.[15] Similarly, GPR have shown a great performance in complex potential energy surface fittings,[33-36] predicting spectroscopic constants of diatomic molecules[37, 38] and second virial coefficients of organic and inorganic compounds.[39] The selected SQM model was PM7,[17] which is overall the most accurate method implemented in MOPAC2016.[40] To fully exploit the SQM calculations, several input features are constructed from the electronic and structural properties of reactant, transition states and products. Unlike some of the previous ML models, entropic effects can be approximately computed from the available structure and vibrational frequencies of both reactant and transition states. This important feature of our model enables the prediction of rate constants at different temperatures. Moreover, the model makes different predictions for cases where two transition states exist for the same rearrangement.[41] A similar synergistic SQM/ML approach to predict activation energies for a diverse class of C-C bond forming nitro-Michael additions has been recently proposed.[42]

The ML correction proposed in this paper could be employed in rapid screening of the large reaction networks that appear in Combustion Chemistry or in Astrochemistry. SQM-based methods for automated reaction mechanism prediction like AutoMeKin[43-46] could also greatly benefit from this machine learning model.

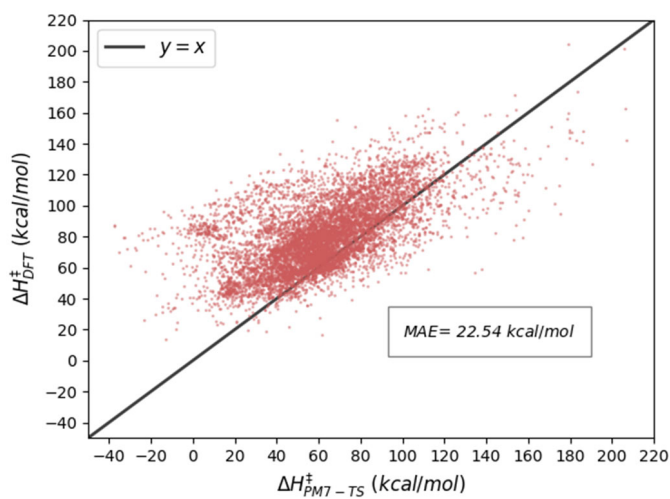

**Fig. 1** Performance of PM7-TS on the GPOC data set.

## 2. Methods

### 2.1 Performance of PM7-TS on the GPOC data set.

Since the accuracy of PM7-TS is uncertain (*vide supra*), its performance was evaluated on the GPOC data set of barrier heights.[7] Figure 1 shows the correlation between the ωB97X-D3/def2-TZVP barrier heights and the values predicted by PM7-TS. In general, PM7-TS significantly underestimates the BHs with a MAE of 22.5 kcal/mol, which is in line with the deviation obtained by Jensen and co-workers on a data set of model enzymes[23] and much greater than the reported error of 3.8 kcal/mol on the training set employed to optimize the PM7-TS parameters.[17] By contrast, the MAEs of a different

SQM method (AM1) evaluated on a dataset of 1000 Michael addition reactions was only 5.71 kcal/mol, with SQM overestimating in general the barrier heights,[42] although in this case all BHs were below 45 kcal/mol.

These results call for an alternative method to predict accurate SQM-based BHs. The proposal of the present work is to employ machine learning models to correct the SQM values.

**2.2 Data set curation.**

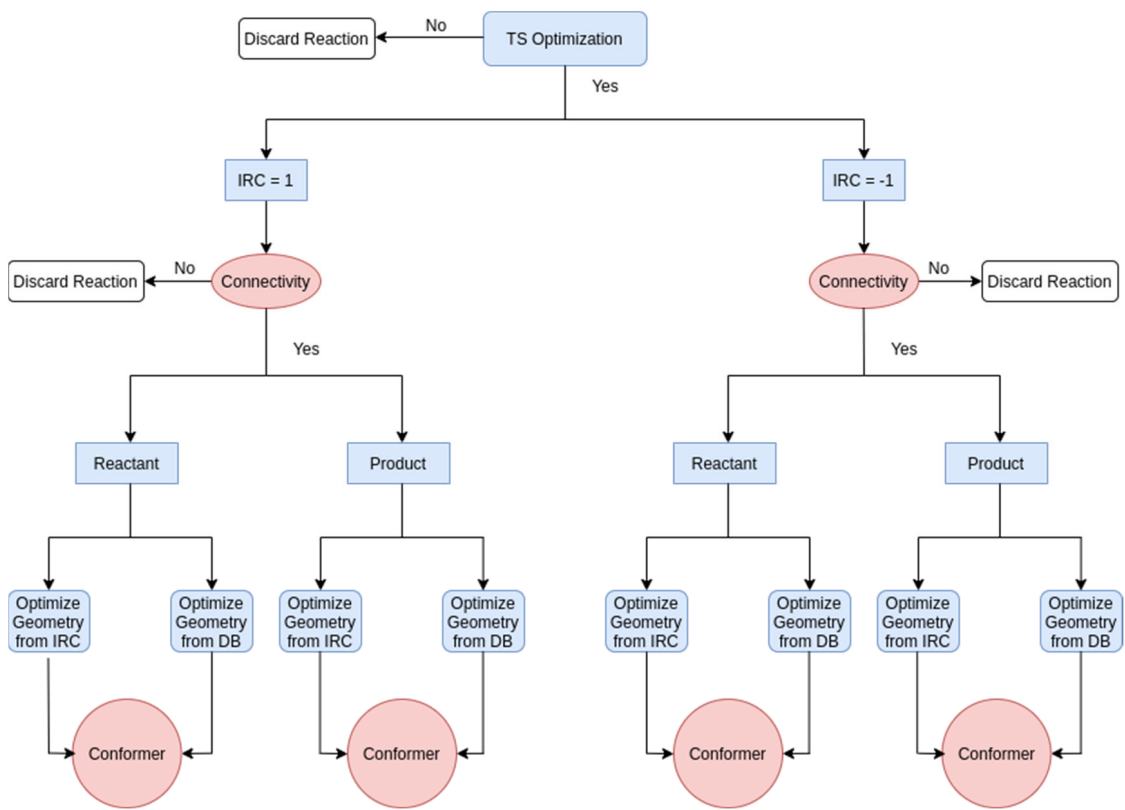

**Fig. 2** SQM data set generation flow diagram.

The target for our machine learning models is the difference $BH^{DFT} - BH^{PM7}$, where $BH^{DFT}$ and $BH^{PM7}$ are the barrier heights obtained at the benchmark (DFT) and PM7 levels, respectively. The GPOC dataset developed by Green and co-workers is employed here.[7] It contains 11,960 reactions, with energies for reactant, transition states and product obtained at the ωB97X-D3/def2-TZVP level of DFT theory. The DFT BHs were directly obtained by subtracting the reactant energy from the TS energy including their ZPEs. Obtaining BHs at the PM7 SQM level entails a more involved process, as several sanity checks are required. A flow chart diagram explaining how the PM7 BHs were obtained is shown in Figure 2. The first step, labeled as TS optimization in the figure, consists of optimizing the TSs at the PM7 level using as initial guesses the geometries optimized at the DFT level. Some structures could not be optimized at the PM7 level and were discarded. Then, for each successfully optimized TS structure, an IRC calculation[47] is carried out in each direction (IRC=1 and IRC=-1 in the figure). The IRC end points are compared with the reactant and product present in the dataset. For such a comparison, the eigenvalues of the corresponding adjacency matrices (with their diagonals representing the atomic numbers) were employed.[48] Obtaining identical eigenvalues ensures that the connectivity of each structure (reactant and product) is the same at both levels of theory of theory (DFT and SQM). When the connectivity differs for either reactant or product, the reaction is discarded. Otherwise, both the IRC end point and the structure from the dataset are optimized at the PM7 level and compared to ensure they present the same conformation. For this last comparison that involves 3D structures, the eigenvalues of a weighted adjacency matrix are employed.[48]

From the initial 11,860 reactions, 8,355 survived this screening process, meaning that roughly 70% of the samples could be utilized in our model. The use of reverse barrier heights did not lead to a major improvement during training of the model but increased the computational cost, so this form of data augmentation was discarded.

An exploratory data analysis (EDA) of the curated GPOC dataset was then carried out. The detailed results of our EDA are collected in the Electronic Supplementary Information (ESI). Reactions in the dataset contain up to seven heavy atoms (C, N or O) per molecule and consist of unimolecular reactions leading to one or more products (although most reactions are isomerizations).

**2.2 Machine learning models.**

Figure 3 shows the workflow for the two machine learning models employed in this study to correct SQM barrier heights. A crucial step of the models is the calculation of a set of descriptors (or input features) that encode the most useful information present in every reaction. Our models employ two types of descriptors: a) standard RDKit-based descriptors and b) a custom set based on the SQM calculations and chemical intuition. Figure 3 also shows how every species in the reaction (namely TS, reactant, or product) contributes to each set of descriptors.

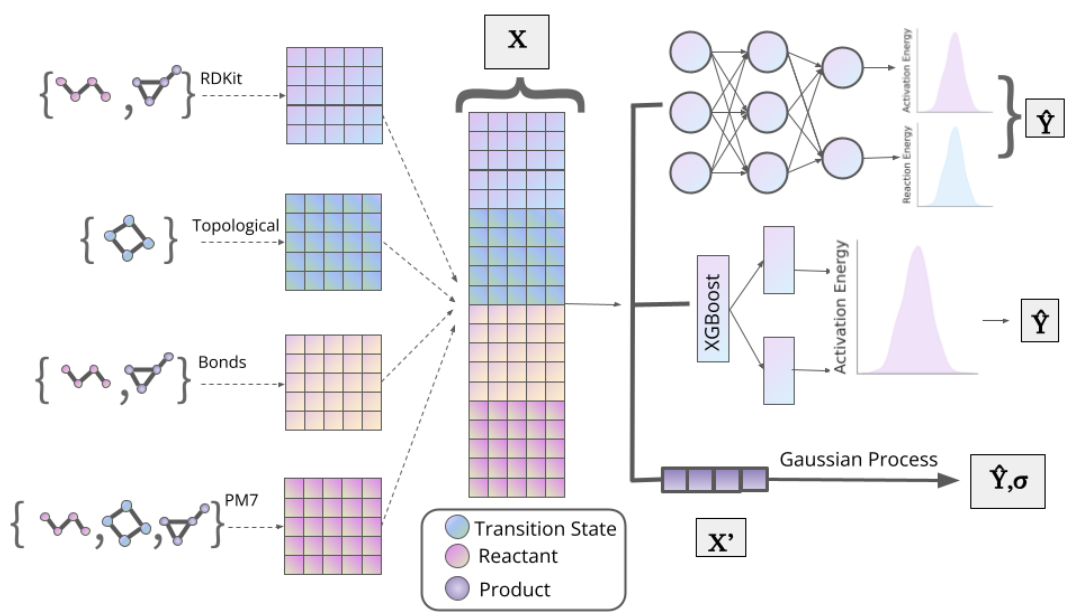

**Fig. 3** Workflow for the prediction of barrier heights using three machine learning models to correct SQM barrier heights. Two types of descriptors are employed: standard RDKit-based and our own custom set that comprises three subtypes. These features **X** are input to DNN and XGBoost regressors, whereas the input features for the GPR are labeled by **X'**. The DNN model predicts the BH and the energy difference between reactant and product. The XGBoost and GPR models predict the BH only.

The first set of descriptors $X_{RDKit}$ is obtained from the cheminformatics library RDKit.[49] Each descriptor of this type $X_{RDKit,i}$ is calculated as:

$$X_{RDKit,i} = X^P_{RDKit,i} - X^R_{RDKit,i} \qquad (1)$$

where $X^R_{RDKit,i}$ and $X^P_{RDKit,i}$ refer to the *i-th* RDKit descriptor of the reactant and product, respectively. If a specific descriptor remains invariant in the reaction (like the molecular weight), the raw value is employed instead of eq 1. The $X_{RDKit}$ set contains 132 descriptors (see the ESI for details).

Besides the above standard set of descriptors, a custom set is also employed in this work. This set is specifically tailored to extract the most relevant features of chemical reactions. It comprises information on the topology of the TS, number of bonds that change in the reaction, and results from the PM7 calculations.

An advantage of our model is that the approximate TS structures calculated at the PM7 level of theory can be employed as input features. Specifically, the 3D geometries are converted into molecular graphs, represented in the form of adjacency matrices, using the definitions employed in AutoMeKin.[46] From the molecular graphs, some topological descriptors $X_{topol}$ can be constructed. These include Randic's connectivity index,[50] the spectral gap (or lowest non-zero eigenvalue of the Laplacian matrix, $\lambda_1^{TS}$), the Estrada index,[51] or the Zagreb index;[52] the full list of topological descriptors can be found in the ESI. The Laplacian matrix defined as $D - A$ (with $D$ and $A$ being the degree and adjacency matrices, respectively) is calculated from a weighted adjacency matrix to account for 3D structures of the TSs.[46] Topological descriptors provide a measure of the extent of branching, or the tightness of the TS structure.

The subset $X_{bonds}$ includes the number of broken and formed bonds of each type, *i.e.*, all pairings of H, C, N and O atoms. For instance, this set includes the descriptors +CO and −CH, which refer to the number of formed CO bonds and number of broken CH bonds, respectively.

The last subset of descriptors $X_{PM7}$ capitalizes on the PM7 calculations. This set includes, the barrier height BH, a rough proxy for the rate constant $e^{-BH}$, the imaginary frequency at the TS $v_1^{TS}$ and differences between ZPEs of reactant, product and TS: $ZPE^R$, $ZPE^P$, $ZPE^{TS}$, respectively. The subset also comprises electronic descriptors like the eigenvalues of the bond order matrix calculated at the TS, the global "hardness"[53] and Mulliken's electronegativity[54] at the TS ($\eta^{TS}$ and $\alpha^{TS}$, respectively), and differences between the self-polarizability[55] of reactant $\pi_R^S$ and product $\pi_P^S$. While some of these descriptors are readily available from a frequency calculation at the TS, others are obtained using the keyword SUPER in MOPAC.

Having defined the input features, a correlation matrix was built where each entry represents the Pearson coefficient $r$ for every pair of descriptors. A threshold was established such that if the correlation coefficient exceeds this value, one of the descriptors is dropped from the input features. The threshold was optimized by cross-validation and set to $r = 0.9$.

These stacked descriptors are input to the three models depicted in Figure 3: DNN, XGBoost and GPR. DNN works in a multi-task approach, where the output includes, besides the barrier height, the energy difference between reactant and product. This approach has shown to enhance predictions and generalization power, even if our interest is only on the BHs.[6, 56] The architecture of the DNN model (number of hidden layers and number of neurons) as well as other hyperparameters were fine-tuned in a 5-fold cross-validation fashion using a grid search, considering some hyperparameters to be orthogonal.

Nevertheless, since our set of descriptors consists of heterogeneous tabular data and the amount of data is limited by deep learning standards, we decided to use two alternative approaches, that perform better in this case, XGBoost and GPR.[57] In particular, we chose

XGBoost[32] implementation of gradient boosting techniques, which achieves SOTA results and provides sparsity-aware algorithms particularly suited for our dataset. In this case, hyperparameters were optimized by means of the Bayesian optimization library Optuna,[58] using 5-fold cross validation as well. Furthermore, considering that Gradient boosting techniques that rely on decision trees as weak learners assign higher importance to descriptors that will be more relevant for other models, we find a more succinct descriptor X' containing only 49 features to feed in the GPR model. The GPR model, after being exposed to the training data generates a multivariate Gaussian prior distribution that by means of Bayesian inference leads to a posterior distribution for the test set. Thus, leading to a prediction with a confidence interval based on the inherent Bayesian nature of the model.

The obtained ML corrections (with DNN, XGBoost or GPR) to PM7 will be referred to as PM7+ML.

## 3. Results and discussion

### 3.1 Performance of the machine learning models.

Following common practices in the ML literature as well as considering the size of the data set, the data was split into 85% training, 5% validation and 10% test. Figure 4 shows the correlation between the reference (DFT) vs the predicted values of the BHs obtained with the XGBoost (XGB) and GPR models for the test set. The MAEs for PM7+ML using as ML models DNN, XGB and GPR are 3.69 kcal/mol, 3.39 kcal/mol and 3.57 kcal/mol, respectively. The performance of PM7+ML is slightly better than Green's for the same number of training points and way better than either PM7 or PM7-TS. It should be noted here that the procedure employed in this work cannot exploit the sort of data augmentation employed in Green's work by including reverse reactions, because many of our features refer to the TSs structures which are common to both direct and reverse reactions.

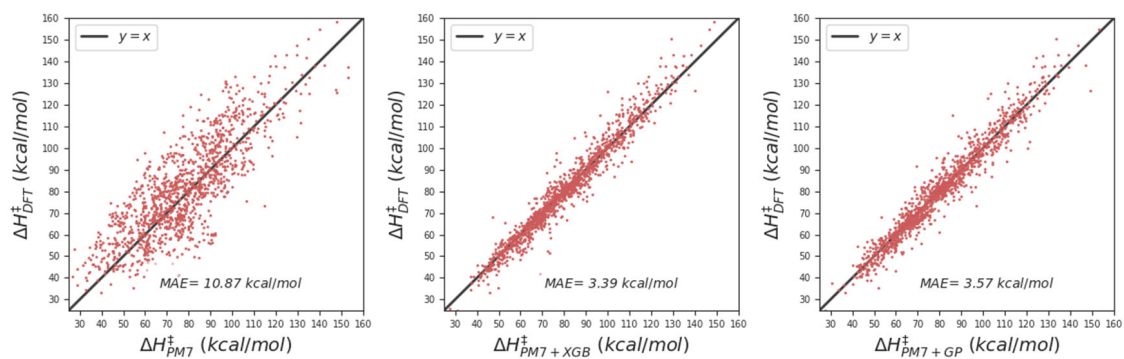

**Fig 4.** Barrier height predictions at DFT, PM7, PM7+XGB and PM7+GP levels.

Figure 5 shows the error distribution on the target variable for XGB and GP in comparison with the results obtained with MOPAC's PM7 calculations. The error distribution for XGB and GP is symmetric, implying that both ML approaches do not suffer from underfitting or overfitting. Besides, the distribution for PM7 is wider than ML models, as expected based on our analysis above on its performance.

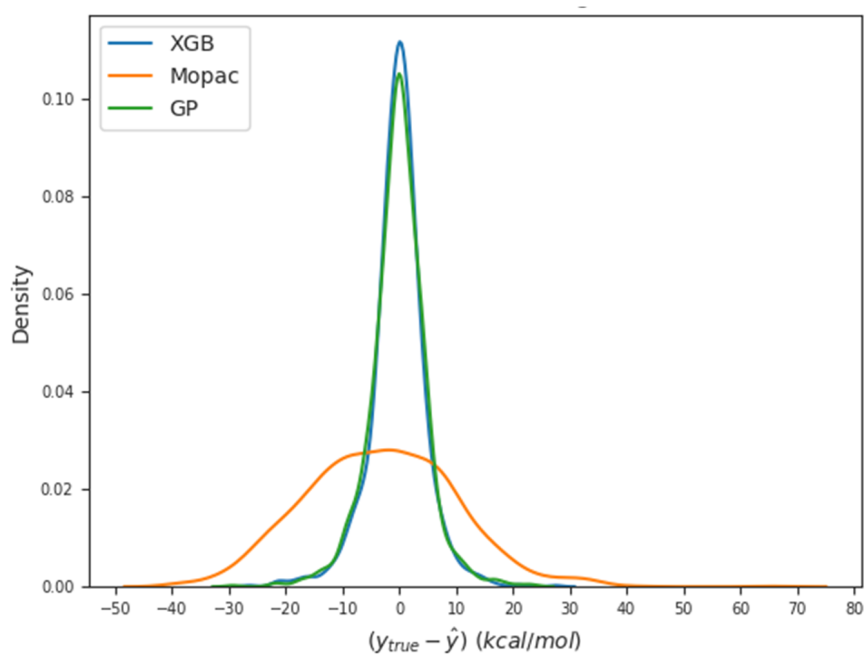

**Fig 5.** Error distribution on the target variable for the ML models (XGB and GP) in comparison with the one obtained directly from the MOPAC calculations.

### 3.2 Interpretability.

Models can be interpreted in terms of their feature importances, i.e., how much a certain feature contributes to the prediction. Feature importances are obtained in present work from the SHAP values,[59] which resort to game-theoretic approaches to measure the contribution to the model output by each descriptor. The underlying principle is to measure the expected change in output when using different combinations of descriptors.

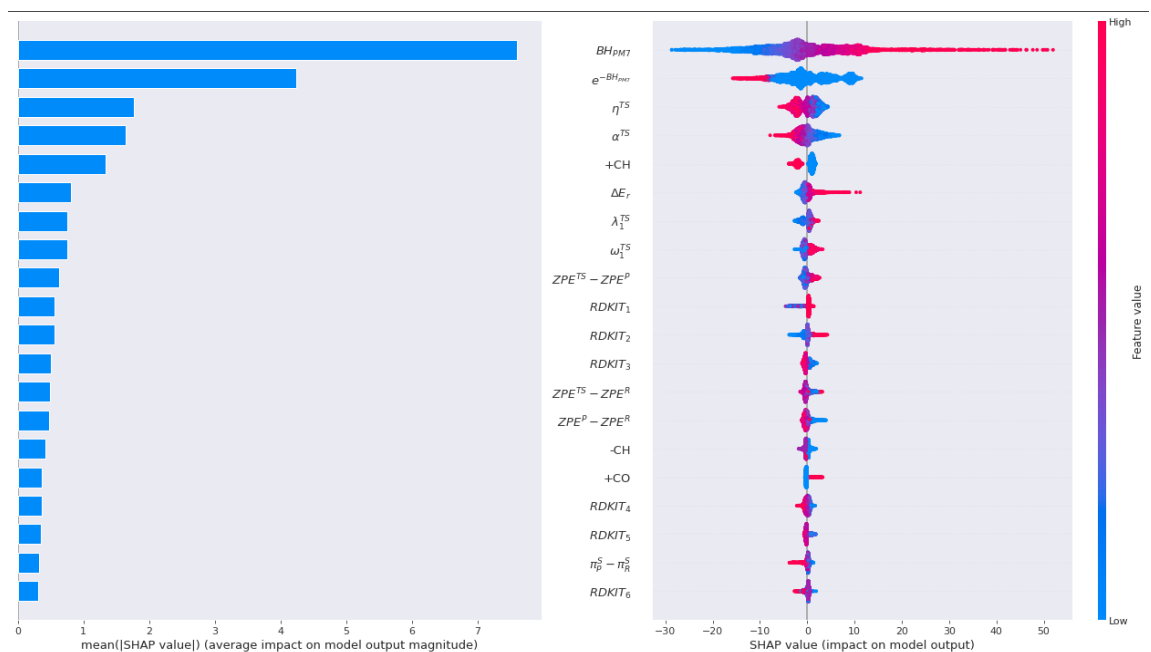

**Fig. 6.** SHAP values for the top 20 most relevant descriptors and their impact on model output.

Figure 6 shows a SHAP summary plot, which displays the magnitude and direction of a feature's effect. Interestingly, 70% (14/20) of the most important features belong to the custom set. As expected, the features with the greatest impact on the model output are the values of the PM7 barrier height $BH_{PM7}$ and the proxy for the rate constant $e^{-BH_{PM7}}$. The figure shows that PM7 tends to underestimate high BHs and viceversa, which is reflected by the positive impact on the model output for high BHs. Our result is in agreement with a recent ML model to predict activation energies from DFT calculations, where the DFT-computed activation energy was also the most important feature.[10]

The "hardness" $\eta^{TS}$ and Mulliken's electronegativity $\alpha^{TS}$ calculated at the TS also rank very high on the global feature importance plot. Using Koopman's theorem[60] they can be approximated as $\eta = (\varepsilon_{LUMO} - \varepsilon_{HOMO})/2$ and $\alpha = -(\varepsilon_{LUMO} + \varepsilon_{HOMO})/2$, where $\varepsilon_{LUMO}$ and $\varepsilon_{HOMO}$ are the energies of the lowest unoccupied molecular orbital (LUMO) and of the highest occupied molecular orbital (HOMO), respectively. These descriptors have been employed as an index to predict chemical behavior and reactivity[61-67] and even to locate TSs.[68] The value of $\eta$ decreases as the molecule departs from its equilibrium position, attaining a minimum at the TS. The LUMO/HOMO energies have also been employed to predict activation energies in Diels-Alder reactions.[11]

With similar impacts on the model output, the absolute value of the imaginary frequency $\omega_1^{TS}$ and the lowest non-zero eigenvalue of the Laplacian $\lambda_1^{TS}$ (or spectral gap) at the TS are also among the most important descriptors according to Figure 6. Both provide a measure for the tightness of the TS structure, with the imaginary frequency also containing information on the mass of the atoms involved in the reaction coordinate.

Number of formed CH and CO bonds (+CH and +CO, respectively), number of broken CH bonds (–CH), ZPE differences among reactant, TS and product, the PM7 reaction enthalpy ($\Delta E_r$) or the self-polarizabilities ($\pi_R^S$ and $\pi_P^S$) also contribute among the most important features. The importance of the number of formed/broken bonds of different types in the model output can be explained by the accuracy of SQM methods predicting bond energies, which strongly depends on the bond type.[69]

RDKIT descriptors considered important include SMR_VSA (RDKIT 1), LabuteASA (RDKIT 2)[70] and VSA_EState2 (RDKIT 7). These descriptors grant a measure of the approximate accessible van der Waals surface area per atom. Other relevant descriptors are: Balaban J (RDKIT 3), referring to the connectivity distance of the molecular graph,[71] Chi0_v[72] (RDKIT 6), which is also a topological based descriptor, and MolLogP[73] (RDKIT 5), which refers to atom-based partition coefficients.

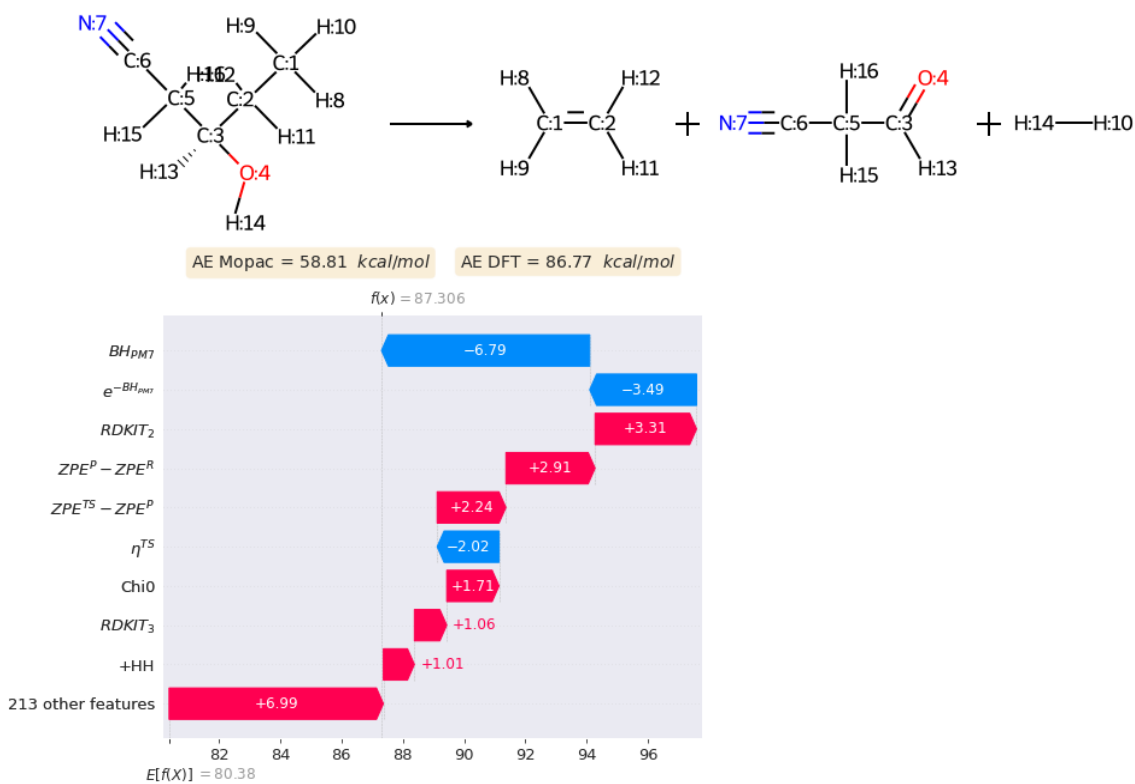

**Fig 7.** Model output interpretation for a single reaction.

Figure 7 showcases how SHAP values can be used for interpretation of a single reaction. It shows the descriptors that contribute the most to shift the prediction of the model from its average (expected) prediction. Not surprisingly, $BH_{PM7}$ and $e^{-BH_{PM7}}$ as well as other descriptors of Figure 6 contribute significantly also for this particular reaction. Additionally, since this reaction involves formation of molecular hydrogen, the number of formed H-H bonds (+HH) is also an important descriptor.

**3.3 Entropic effects.**

In mechanistic and kinetics studies of chemical reactions the quantity of interest is the Gibbs energy of activation $\Delta G^\ddagger$, rather than the BH. The reason is because the former includes enthalpic and entropic corrections to the electronic and ZPE energies. Reaction channels that are not very competitive at low temperatures/energies might become predominant at high temperatures/energies because of the entropic factors.[74] Therefore, the prediction of $\Delta G^\ddagger$ is crucial when the interest is the kinetics and the determination of the predominant mechanism.

The calculation of $\Delta G^\ddagger$ is straightforward when the geometries and vibrational frequencies of the reactant and TS are available. An advantage of our model is that this data is available, though at the approximate SQM level of theory. The values of $\Delta G^\ddagger$ have been obtained in this work at different temperatures using the thermochemistry module of AutoMeKin[46] for the reference and SQM calculations using the rigid rotor/harmonic oscillator approximation. In the absence of a scaling factor for the ωB97X-D3/def2-TZVP vibrational frequencies, the value of 0.9914 was employed; this is the recommended value for the related ωB97X-D/def2-TZVP model chemistry.[75] Furthermore, the PM7 vibrational frequencies were corrected using the recommended scaling factors.[76]

Figure 8 displays the correlation between the reference (DFT), the PM7, and the PM7+ML prediction for $\Delta G^\ddagger$ at three different temperatures: 300, 500 and 1000 K. At the two lowest temperatures, the PM7+ML prediction is roughly of the same accuracy as that for the BH. However, for the highest temperature of 1000 K, the ML prediction starts to deteriorate and the MAE at this temperature is 5.30 kcal/mol. A clear improvement to the model would be to use a multitask ML model to correct the SQM vibrational frequencies.

Nevertheless, the current accuracy of the PM7+ML model significantly improves on the PM7 accuracy, and it may suffice for fast screening of reaction networks.

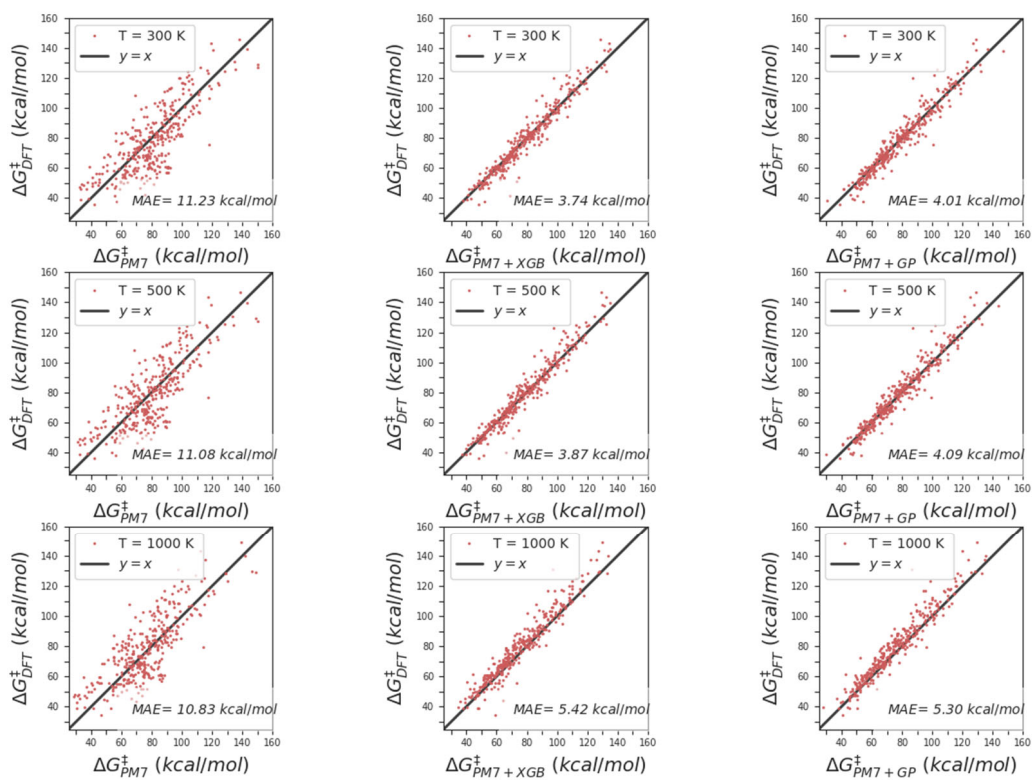

**Fig 8.** Correlation of the DFT, PM7, PM7+XGB and PM7+GP values for the Gibbs energy difference between the reactant and transition state $\Delta G^{\ddagger}$ at $T = 300$, 500 and 1000 K.

## 4. Conclusions

The main conclusions of this work are summarized below:

a) Cheap SQM calculations can be leveraged to obtain DFT-quality barrier heights by means of ML.

b) The MAEs of our ML models (multi-task deep neural network, gradient boosted trees by means of the XGBoost interface, and Gaussian process regression) are of the same magnitude as those obtained in previous work.

c) The analysis of the models shows that the custom-made descriptors obtained from the MOPAC calculations are, in general, considered more important than those obtained from standard cheminformatics libraries.

   d) Our MOPAC-based descriptors could be widely adopted in future quantitative predictions of reaction properties.

   e) An additional advantage of our ML models is the inclusion of entropic effects, which is very important for predicting the kinetics at different temperatures.

   f) Our ML models could be used for screening large reaction networks, or they could be implemented in automated reaction mechanism programs based on SQM calculations.

**Conflicts of interest**

There are no conflicts of interest to declare.

**Acknowledgements**

This work was partially supported by Consellería de Cultura, Educación e Ordenación Universitaria (Grupo de referencia competitiva ED431C 2021/40), and by Ministerio de Ciencia e Innovación through Grant #PID2019-107307RB-I00.

# Supporting Information for:

# Barrier height prediction by machine learning correction of semiempirical calculations

Xabier García-Andrade,[a,b] Pablo García Tahoces[b] , Jesús Pérez-Ríos[c,d] and Emilio Martínez Núñez*[a]

[a]Department of Physical Chemistry, University of Santiago de Compostela 15782, Spain
[b]Department of Electronics and Computer Science, University of Santiago de Compostela 15782, Spain
[c]Department of Physics, Stony Brook University, Stony Brook, New York 11794, USA
[d]Institute for Advanced Computational Science, Stony Brook University, Stony Brook, NY 11794-3800, USA

# Contents

# 1.Exploratory Data Analysis:

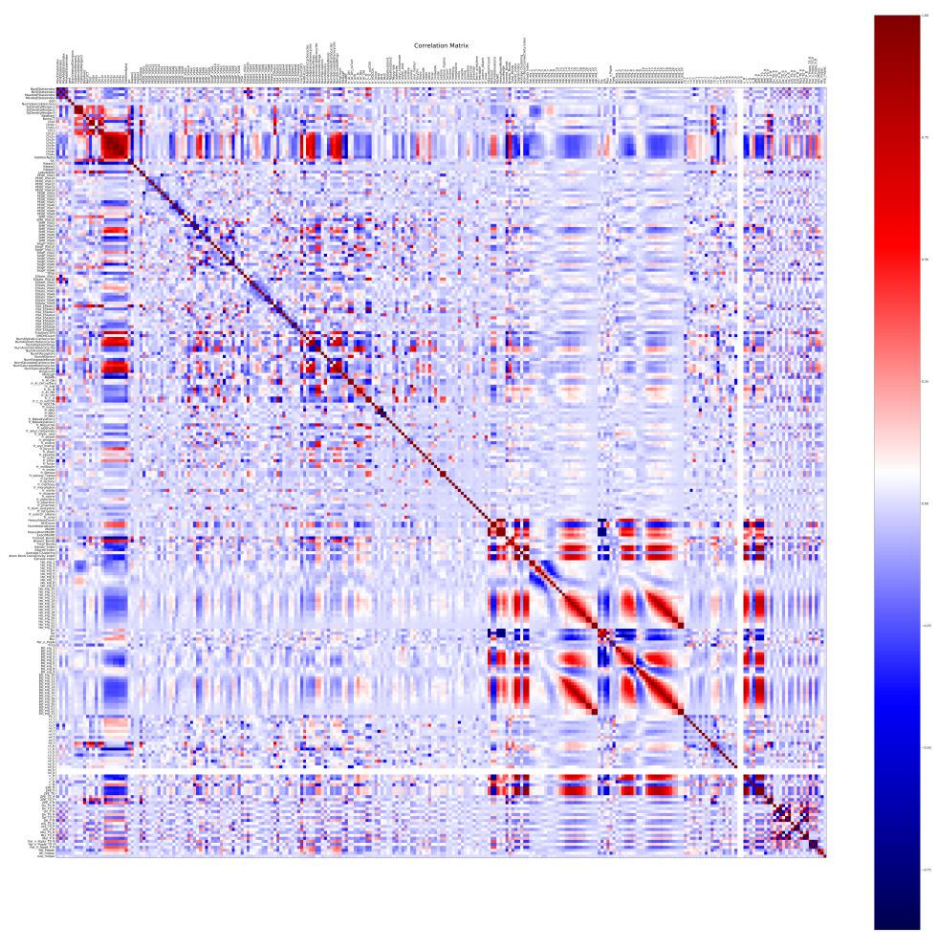

**Fig S1**. Correlation matrix, computing the pairwise Pearson coefficient for each pair of descriptors.

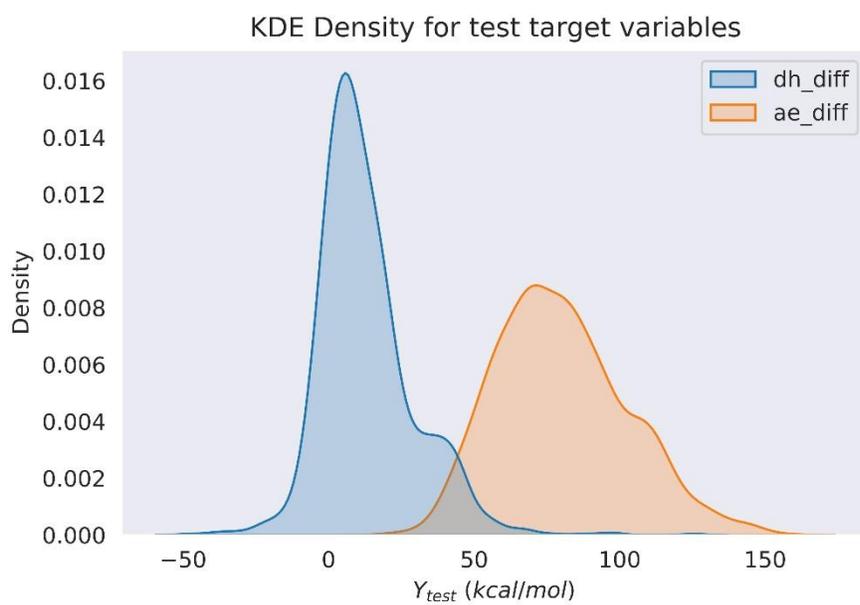

**Fig S2**. Test set target variables distribution.

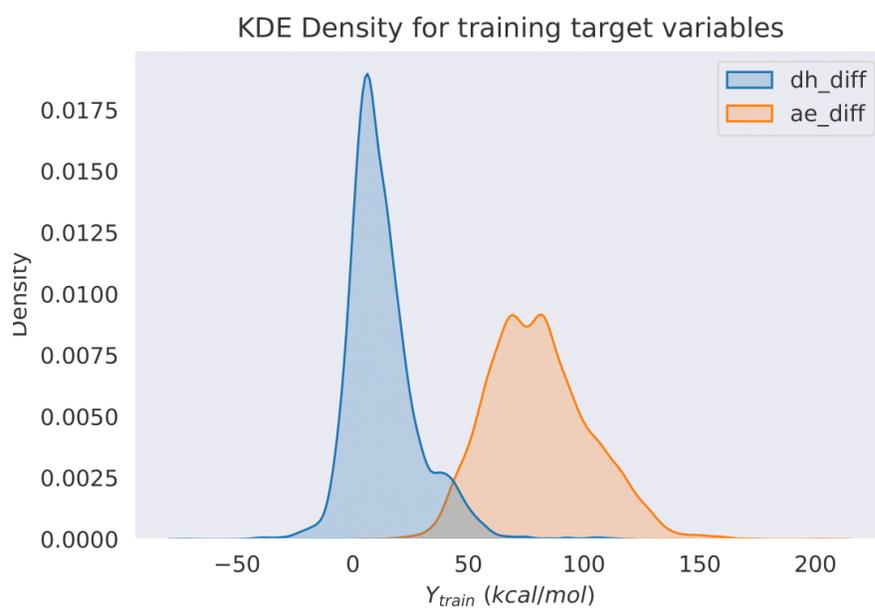

**Fig S3**. Train set target variables distribution.

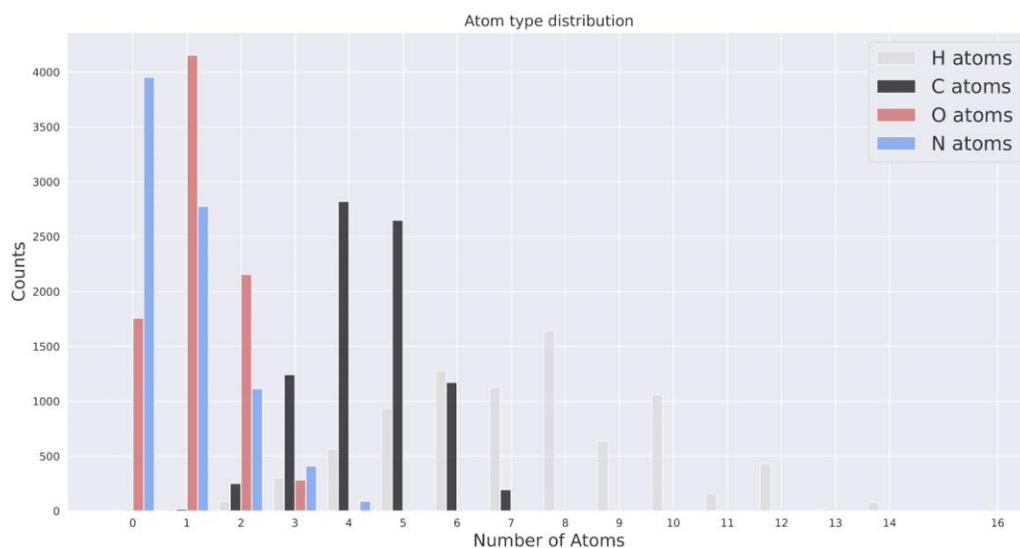

**Fig S4.** Type of atom count per reaction in the whole database.

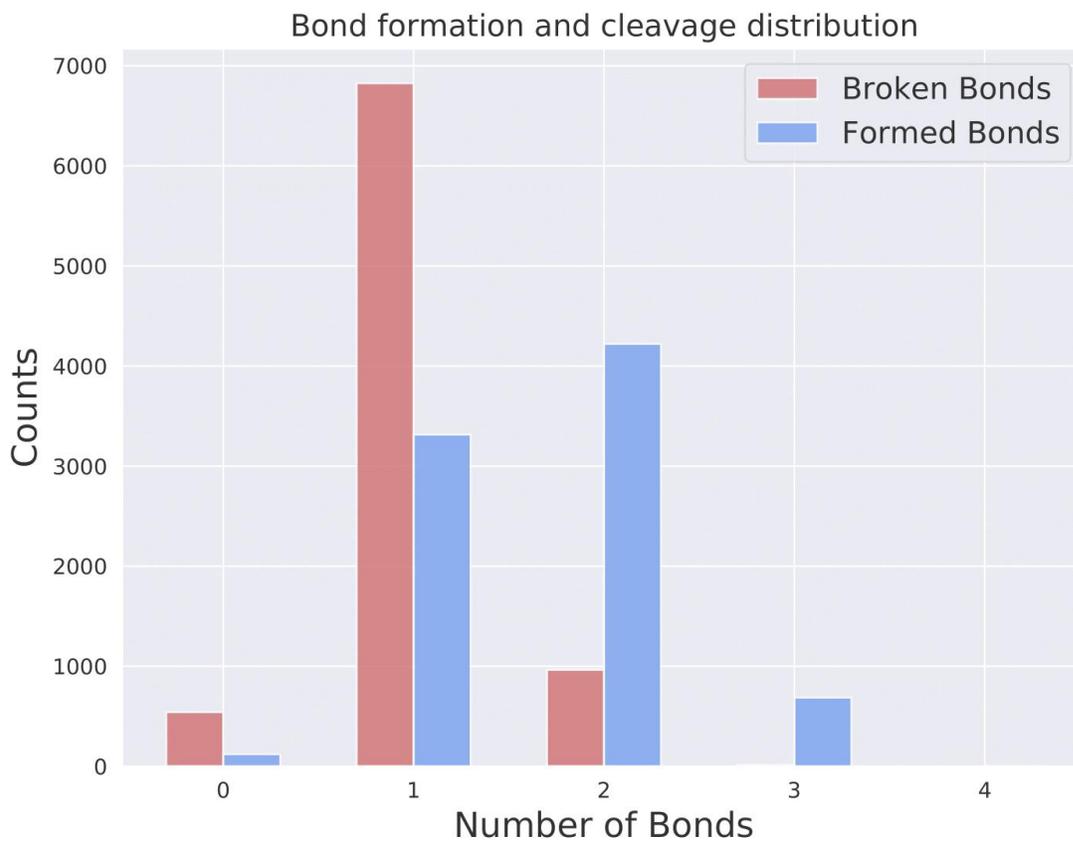

**Fig S5.** Frequency of the number of broken and formed bonds per reaction.

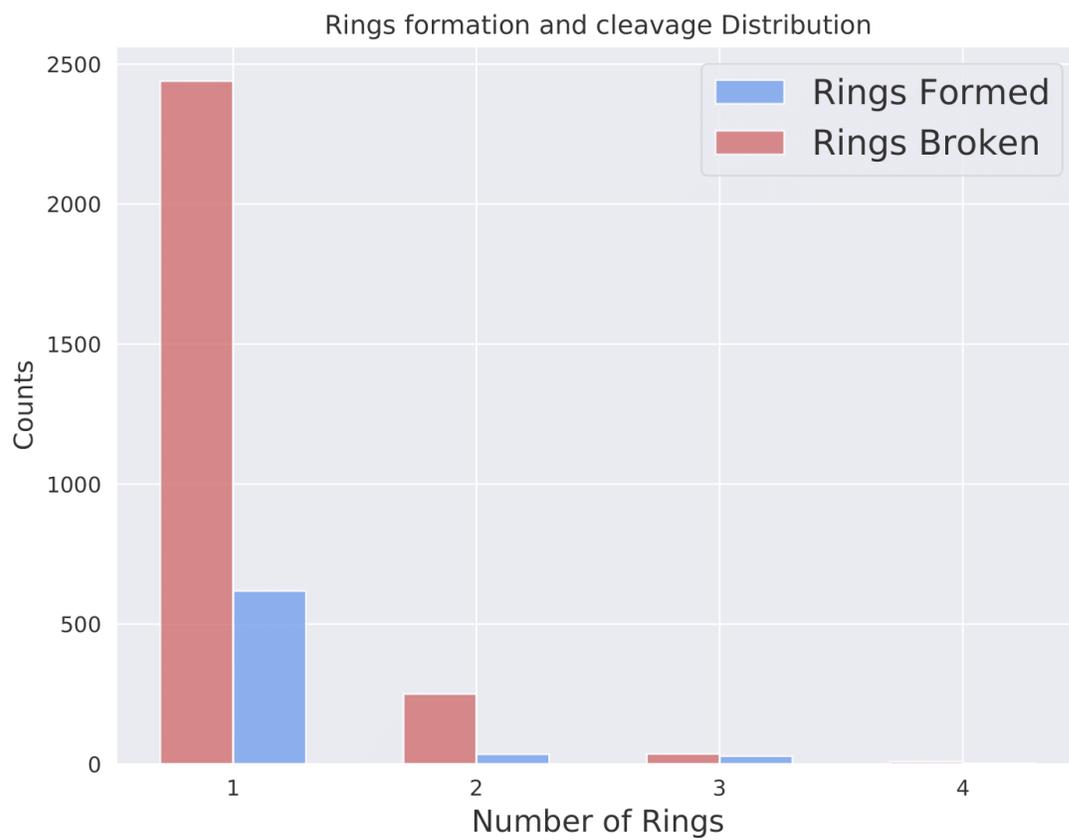

**Fig S6.** Frequency of ring formation and ring cleavage per reaction.

## 2. Hyperparameter Optimization:

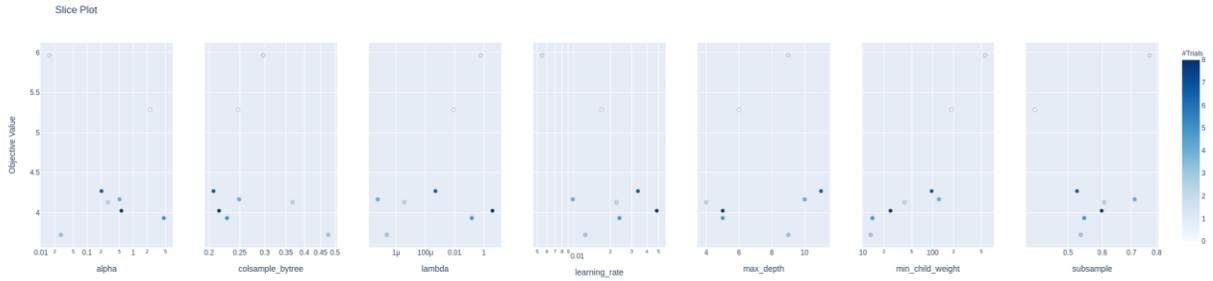

**Fig S7.** Optuna values per trial in the hyperparameter optimization iterations.

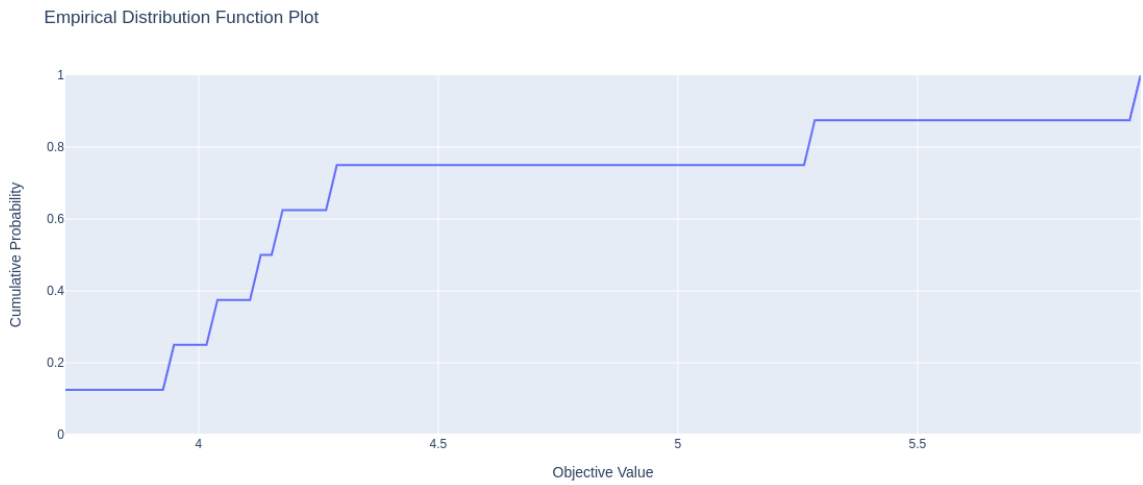

**Fig S8**. Optuna empirical cumulative probability.

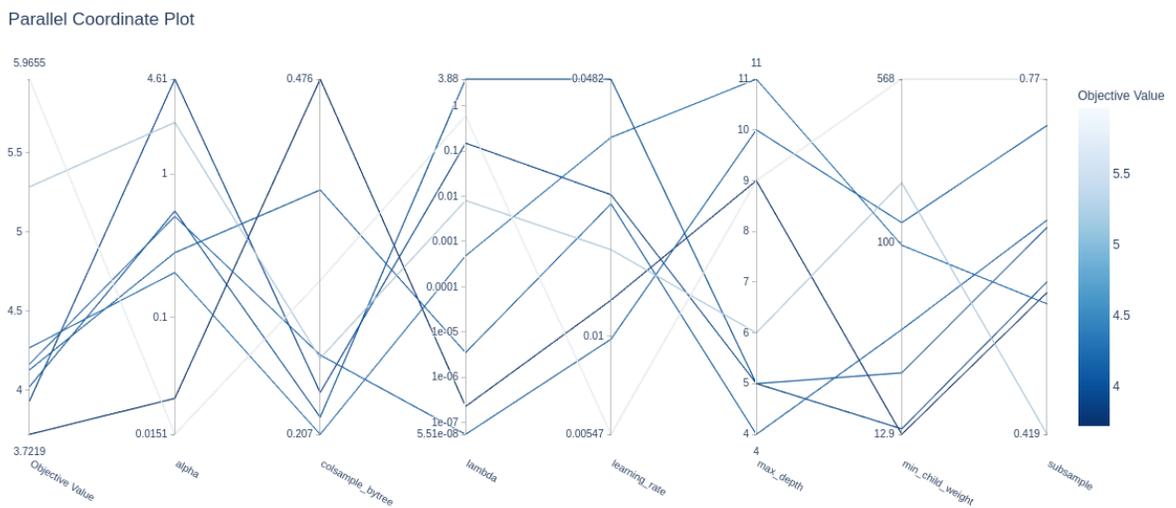

**Fig S9**. Interactions between the hyperparameters and objective value .

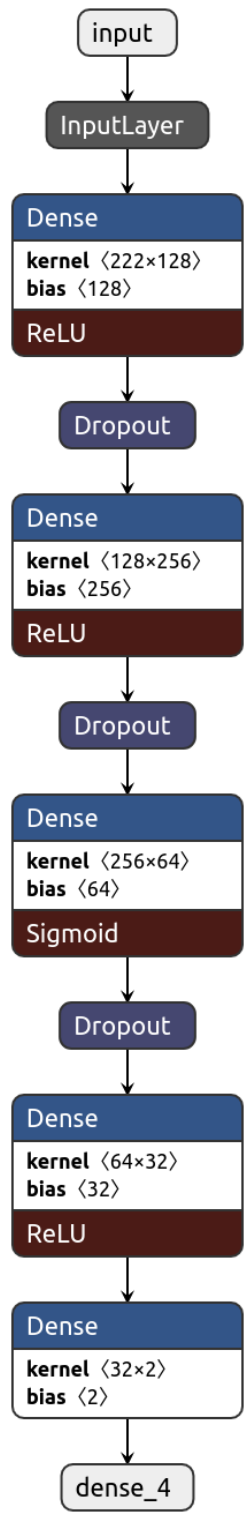

**Fig. S10.** DNN architecture after hyperparameter tuning.

# 3. Descriptor explanation:

List of descriptors in the dataset (using the same name convention as in the provided dataset):

- Graph-theoretic (topological) descriptors:
    - randic_index: captures the contributions per bond to each vertex. [1]
    - zagreb_index: graph invariant descriptor [2].
    - avg_clustering: descriptor that measures how close an atom is to its neighbors. [3]
    - atom_bond_conn_idx: provides an estimate to the strain energy of cycloalkanes. [4]
    - estrada_index:
    - lap_eig: Eigenvalues of the transition state Laplacian matrix, which need to be padded to the maximum number of atoms in the dataset.
- MOPAC descriptors:
    - Dn, De, piS, Mul, Par_n_Pople: Nucleophilic and Electrophilic delocalizabilites, Self-Polarizability, Mulliken Electronegativity, and Parr & Pople absolute hardness. These are quantities derived from the density matrix [5].
    - Freq: first real-valued frequency of a transition state.
    - BO_eigs: Eigenvalues of the bond order matrix provided by the MOPAC calculation.
    - ZPE: Zero Point Energy computed using MOPAC.
    - Differences of the previous quantities between the entities involved in the reaction (e.g. ZPE_TS_R refers to subtracting the ZPE of the reactant from the ZPE of the transition state.
    - AE_mopac and exp_mopac: barrier height and minus the exponential of the barrier height computed at MOPAC level of theory.
    - DH_mopac: reaction enthalpy computed at MOPAC level of theory.
- Bond descriptors:
    - Broken_bonds: Number of broken bonds.
    - Formed_bonds: Number of formed bonds.
    - Total_bonds: Number of total bonds.
    - Counts of the type of atoms in the reaction (e.g. n_H is the count of Hydrogen atoms in the reaction).
    - Counts of the type of bonds involved in the reaction (e.g. co_f refers to the number of C-O bonds formed in the reaction).
- RDKit descriptors [6].

The data and code are available in the following repository:
https://github.com/XabierGA/SemiDFT